\documentclass[conference]{IEEEtran}
\IEEEoverridecommandlockouts
\usepackage{cite}
\usepackage{amsmath,amssymb,amsfonts}
\usepackage{algorithmic}
\usepackage{graphicx}
\usepackage{textcomp}
\usepackage{xcolor}

\usepackage{adjustbox}
\usepackage{amsthm}
\usepackage{color}
\usepackage{enumitem}
\usepackage{tabu}
\usepackage{tikz}
\usetikzlibrary{matrix,shapes,arrows,positioning,chains, calc, automata}
\usepackage[]{todonotes} 
\usepackage{subcaption}
\newcommand\copyrightnotice[1]{
	\begin{tikzpicture}[remember picture,overlay]
		\node[anchor=south,yshift=10pt] at (current page.south) {\fbox{\parbox{\dimexpr\textwidth-\fboxsep-\fboxrule\relax}{#1}}};
	\end{tikzpicture}
}

\theoremstyle{plain}
\newtheorem{thm}{Theorem}

\theoremstyle{definition}

\theoremstyle{definition}
\newtheorem{property}[thm]{Security Property}

\tikzset{every tree node/.style={align=center}}

\DeclareMathAlphabet{\pazocal}{OMS}{zplm}{m}{n}

\newcommand*{\customParagraph}[1]{
	\medskip 
	
	\noindent \textbf{#1}}

\newcommand*{\framework}{framework }
\newcommand*{\secprop}{Security Property}

\newcommand*{\Pseudonym}[2]{\ensuremath{\Phi_{#1}^{#2}}}
\newcommand*{\transcript}[1]{\ensuremath{\mathcal{T}(#1)}}
\newcommand*{\prob}[1]{\ensuremath{\textnormal{Pr}(#1)}}

\newcommand*{\auth}[1]{\ensuremath{\textnormal{Auth}(#1)}}
\newcommand*{\rnd}[1]{\ensuremath{\textnormal{Rnd}(#1)}}
\newcommand*{\sig}[1]{\ensuremath{\textnormal{Sig}(#1)}}
\newcommand{\threshold}{\ensuremath{\textnormal{T}}}
\newcommand*{\pubkey}[1]{\ensuremath{\textnormal{pk}_{#1}}}
\newcommand*{\privkey}[1]{\ensuremath{\textnormal{sk}_{#1}}}

\newcommand*{\draw}{\ensuremath{\leftarrow}}

\newcommand*{\partysymbol}[1]{\ensuremath{P_{#1}}}
\newcommand*{\parties}{\ensuremath{\mathcal{P}}}

\newcommand*{\numparties}{\ensuremath{\iota}}

\newcommand*{\numcorparties}{\ensuremath{\kappa}}
\newcommand*{\corparties}{\ensuremath{\mathcal{C}}}

\newcommand*{\func}{\ensuremath{\pazocal{F}}}

\newcommand*{\adversary}{\ensuremath{\mathcal{A}}}

\def\BibTeX{{\rm B\kern-.05em{\sc i\kern-.025em b}\kern-.08em
    T\kern-.1667em\lower.7ex\hbox{E}\kern-.125emX}}
\begin{document}

\title{Introducing a Framework to Enable Anonymous Secure Multi-Party Computation in Practice (Extended Version)\\
\thanks{This work was funded by the Deutsche Forschungsgemeinschaft (DFG, German Resarch Foundation) - project number (419340256) and NSF grant CCF-1646999. Any opinion, findings, and conclusions or recommendations expressed in this material are those of the author(s) and do not necessarily reflect the views of the National Science Foundation.}}

\author{\IEEEauthorblockN{Malte Breuer}
\IEEEauthorblockA{\textit{Department of Computer Science} \\
\textit{RWTH Aachen University}\\
Aachen, Germany \\
breuer@itsec.rwth-aachen.de}
\and
\IEEEauthorblockN{Ulrike Meyer}
\IEEEauthorblockA{\textit{Department of Computer Science} \\
	\textit{RWTH Aachen University}\\
	Aachen, Germany \\
	meyer@itsec.rwth-aachen.de}
\and
\IEEEauthorblockN{Susanne Wetzel}
\IEEEauthorblockA{\textit{Department of Computer Science} \\
\textit{Stevens Institute of Technology}\\
Hoboken, USA \\
swetzel@stevens.edu}
}

\maketitle
\copyrightnotice{\copyright\space 2021 IEEE. This is the author's extended version of the work. It is posted here for your personal use. Not for redistribution. The definitive version was published in \emph{18th Annual International Conference on Privacy, Security and Trust (PST2021), December 13-15, 2021, Auckland, New Zealand}, https://doi.org/10.1109/PST52912.2021.9647793}
\begin{abstract}
	Secure Multi-Party Computation (SMPC) allows a set of parties to securely compute a functionality in a distributed fashion without the need for any trusted external party. Usually, it is assumed that the parties know each other and have already established authenticated channels among each other. However, in practice the parties sometimes must stay anonymous. In this paper, we conceptualize a framework that enables the repeated execution of an SMPC protocol for a given functionality such that the parties can keep their participation in the protocol executions private and at the same time be sure that only authorized parties may take part in a protocol execution. We identify the security properties that an implementation of our framework must meet and introduce a first implementation of the framework that achieves these properties. 

\end{abstract}

\section{Introduction}\label{sec:introduction}

Traditionally, \emph{Secure Multi-Party Computation (SMPC)} allows a fixed set of parties to compute a functionality~$ \func $ in a distributed fashion and without the involvement of a trusted external party. To this end, all parties jointly execute an SMPC protocol such that each party provides a private input and may receive a private output. While there are applications that use SMPC protocols in practice (e.g., the Danish sugar beet auctions~\cite{Bogetoft_SugarBeetAuctions_2009}), such applications typically assume that the parties that participate in the protocol execution know each other and have already established authenticated channels with each other. However, this assumption does not necessarily hold in practice. This is particularly true if the parties must keep their participation in the protocol execution private as this already constitutes sensitive information. For example, consider companies who repeatedly intend to trade certain goods among each other. The frequency at which a company can participate in the computation of a trade may already leak valuable information about the company's production capacity. Thus, a company would want to keep this information private while still seeking assurance that only authorized parties are allowed to participate in the trade computation to avoid trades with unwanted trade partners (e.g., criminal organizations).

To the best of our knowledge, there is no approach to date which allows a party to keep its repeated participation in SMPC protocol executions for computing a certain functionality~$ \func $ private while at the same time guaranteeing that only authorized parties are allowed to participate. It is in this context that this paper provides two main contributions to address this gap. First, we devise a novel \framework that enables the repeated anonymous participation in SMPC protocol executions. Second, we specify a first implementation of the \framework based on the Tor network~\cite{Tor_Project} for anonymous communication and blind signatures for the establishment of authenticated channels among the anonymous parties. 

\subsection{Intuition}

There are two main properties that our framework aims to provide, i.e., ensuring that only authorized parties may participate in the SMPC protocol executions for computing the functionality~$ \func $ and guaranteeing the anonymity of those parties that actually participate in the protocol executions.
To make sure that only authorized parties may participate in the protocol executions, each party that intends to participate in an SMPC protocol execution for the computation of~$ \func $ first has to register with the framework. The set of these \emph{registered parties} then comprises all parties that are considered authorized and are thus allowed to participate in the computation of~$ \func $. 

Since we also want to achieve that the parties' participation in the actual protocol executions for the computation of the functionality~$ \func $ remains private, we have to specify a second set of parties which contains only those parties that currently intend to participate in a protocol execution. This set is called the set of \emph{currently active parties} and it forms a subset of the set of all registered parties. While all registered parties know the overall set of registered parties, it is not known to any of the registered parties which parties are currently active. Thus, the parties that are currently active are anonymous within the set of all registered parties which forms the anonymity set.\footnote{Anonymity in this sense could also be achieved if all registered parties always participated in every protocol execution using dummy inputs to indicate that they currently do not want to compute~$ \func $. However, this would imply~a huge performance overhead as the runtime of SMPC protocols usually increases rapidly with increasing numbers of parties and a party would have to participate in the computation even if it currently does not want to compute~$ \func $.}

Additionally, we require an entity that provides the service (i.e., the possibility to participate in the SMPC protocol executions for the computation of a given functionality~$ \func $) to the parties. We refer to this central entity as the \emph{coordinator}. The coordinator handles the registration of new parties as well as their requests to participate in an SMPC protocol execution for the computation of the functionality~$ \func $. While SMPC in general aims to avoid a central entity, it is inevitable to use a central coordinator here to make sure that only authorized parties may participate in a protocol execution for computing~$ \func $. However, to ensure that parties do not need to simply trust the coordinator, the framework includes a bulletin board which contains information on the coordinator's actions. This bulletin board has to be audited by all registered parties to make sure that the coordinator behaves as intended.

Intuitively, anonymity is achieved using pseudonyms that cannot be linked to the parties' identity. Using their current pseudonym, parties can indicate their intent to participate in a protocol execution for computing the functionality~$ \func $. Such a protocol execution is initiated as soon as a sufficient number of parties have expressed their intent. For the next protocol execution the parties then obtain new pseudonyms that can neither be linked to their identity nor to their previous~pseudonym.

\subsection{Outline}
We first introduce the necessary preliminaries for our work and review the relevant related work (Section~\ref{sec:pre}). In Section~\ref{sec:system_design}, we introduce our novel framework for enabling repeated anonymous SMPC protocol executions together with the considered threat model and formal definitions of the security properties which an implementation of our framework has to meet. Then, in Section~\ref{sec:system_implementation} we present a first implementation of our framework and discuss in how far this implementation meets the previously defined security properties. We conclude with some remarks on future work (Section~\ref{sec:conclusion}).

\section{Background and Related Work}\label{sec:pre}

Our framework combines the two privacy-enhancing technologies secure multi-party computation and anonymous communication. In this section, we briefly introduce the relevant background and related work in these two research areas.

\subsection{Secure Multi-Party Computation}\label{sub:pre_smpc}

In SMPC a set of $ \numparties $ parties securely compute an \linebreak $ \numparties $-input functionality~$ \func: \ ({0,1}^*)^\numparties \rightarrow ({0,1}^*)^\numparties $ such that each party only learns its private input and output. An adversary controlling $ \numcorparties < \numparties $ parties tries to learn as much as possible about the honest parties' input or to manipulate the outcome of the computation. In general, one differentiates between two adversary models. In the \emph{semi-honest} model, the corrupted parties strictly follow the protocol specification and only try to learn as much as possible on the other parties' input. In the \emph{malicious} model, the corrupted parties may additionally arbitrarily deviate from the protocol specification. 

SMPC was first introduced by Yao~\cite{Yao_TwoParty_1986} for the two-party case and by Goldreich et al.~\cite{Goldreich_MentalGame_1987} for the multi-party case. Since then many SMPC protocols have been proposed for different use cases such as secure auctions (e.g.,~\cite{Bogetoft_Auctions_2006}), private information retrieval (e.g.,~\cite{Chor_PIR_1995}), or secure machine learning~(e.g.,~\cite{Liu_SecureML_2017}). 

However, to the best of our knowledge the assumption for any SMPC protocol is always that the parties that execute a protocol know each other and already have authenticated channels established among each other. In contrast to this, our framework allows the parties that intend to repeatedly execute an SMPC protocol for some functionality~$ \func $ to find each other, establish authenticated channels among each other, and keep their participation in the protocol execution private.

\subsection{Anonymous Communication}\label{sub:pre_tor}

To keep their participation in an SMPC protocol execution private, the parties in our \framework have to communicate in an anonymous fashion, i.e., sender and receiver of messages have to remain anonymous towards each other. To date different approaches for anonymous communication have been proposed (e.g., Chaum's Mix Cascades~\cite{Chaum_MixCascades_1981}~or~I2P~\cite{Schimmer_I2P_2009}).

However, the primary choice for low latency anonymous communication is the Tor Network~\cite{Syverson_Tor_2004,Tor_Project}.
The Tor network enables a party to send messages to other parties, keeping the sender identity private. Furthermore, Tor includes the Tor Onion Service Protocol~\cite{Tor_Onion_Service} which allows a party to provide other parties with a so-called \emph{onion URL}. Any other party can send messages to this onion URL without knowing the receiver identity, i.e., the actual identity behind the URL. 
In the implementation of our framework (cf.~Section~\ref{sec:system_implementation}), onion URLs are used as pseudonyms for the parties whenever they require to stay anonymous when receiving a message.

\section{Framework}\label{sec:system_design}

Our framework is geared to allow a set of parties to repeatedly participate in SMPC protocol executions for computing a functionality~$ \func $ such that they can keep their participation in the protocol executions private. 
In the following, we first describe the properties of each of the three main components of our \framework (Section~\ref{sub:components}). Then, we describe the life cycle of a party within the framework (Section~\ref{sub:system_overview}). We conclude the section with a description of the considered threat model (Section~\ref{sub:goals_threatmodel}) and the security properties which an implementation of our framework has to meet (Section~\ref{sub:goals_secGoals}).

\subsection{Components}\label{sub:components}

\customParagraph{Coordinator:} The coordinator is the central entity of the new framework. She is responsible for the registration of all parties that intend to participate in the computation of the functionality~$ \func $ at some point in time. Consequently, she has the power to decide which parties are allowed to participate. The coordinator also handles the requests of already registered parties who intend to participate in an SMPC protocol execution for the computation of~$ \func $. However, the coordinator herself does not participate in the actual protocol execution.

Since the coordinator is the entity who provides the service to the parties, we assume that she in general wants parties to register and use the service. Furthermore, we assume that the coordinator only acts maliciously if it is to her benefit. In particular, she will not act maliciously at random, e.g., she will not deny a party whose actual identity is unknown to her to participate in the computation of the functionality~$ \func $.

\begin{figure*}[t]
	\centering
\begin{adjustbox}{width=\textwidth}
	\begin{tikzpicture}[ 
		> = stealth, % arrow head style
		shorten > = 1pt, % don't touch arrow head to node
		auto,
		node distance = 2cm, % distance between nodes
		semithick % line style
		]
		\tikzstyle{every state}=[
		draw = black,
		thick,
		fill = white,
		inner sep=2pt,
		minimum size = 2.8cm
		]
		
		\draw[] (-2, 3.5) rectangle (23.1, -6.4);
		\node[anchor=west] (system) at (-2, 3.2) {\large Framework};
		
%		\draw[] (23.4, 3.7) rectangle (27, -7.8);
%		\node[rectangle, anchor=west, align=center] (protocol) at (24, 3.2) {\large Outside of\\\large the Framework};
		
		\node[] (s0) at (-2,0) {};
		
		\node[state, align=center, draw] (s1) at (0,0) {new};
		
		\node[state, align=center, draw] (s2) at (6.3,0) {registered,\\only auditing};
		
		\node[state, align=center, draw] (s3) at (12.6,0) {currently active,\\waiting in pool};
		
		\node[state, align=center, draw] (s4) at (18.9, 0) {currently active,\\left pool, exec.\\sanity check};
		
		\node[rectangle, minimum height=3cm, minimum width=3cm, align=center, draw] (s5) at (25.2,0) {SMPC protocol\\execution outside\\of the framework\\(the coordinator\\is not involved)};
	
		%, label={[align=center, color=blue]below:the actual protocol execution\\is not part of the system}
		
		\node[state, align=center, draw] (s6) at (6.3, -4) {registered,\\executing\\sanity check};
		
		\node[state, align=center, draw] (s7) at (17, -4) {malicious,\\banned};
		
		\node[state, align=center, draw] (s8) at (21.4, -4) {all parties,\\stop using\\the service};
		
		%, label={[align=left, color=blue]below right:if a party abandons the system due to\\a malicious coordinator, the party assumes\\that the whole system is corrupted,\\i.e., it will no longer use the system}

		\path[->] (s0) edge (s1);
		
		\path[->] (s1) edge[bend left=0] node[above, align=center] {1. authenticate\\to coordinator} (s2);
		
		%\path[->] (s2) edge[bend left=10] node[below, align=center] {2. revoke registration\\with the coordinator} (s1);
		
		\path[->] (s2) edge[bend left=0] node[sloped, align=center] {2. request to compute\\functionality $ \func $\\using a pseudonym} (s3);
		
		\path[->] (s3) edge[bend left=0] node[above, align=center] {3. pool threshold\\reached} (s4);
		
		\path[->, dashed] (s4) edge node[sloped, align=center, xshift=-0.3cm] {}(s5);
		
		\path[->] (18, 1.2) edge[bend right=20] node[above, align=center] {4. sanity check finished,\\no malicious behavior detected} (7.2, 1.2);
		
		\path[->] (s2) edge[bend left=10] node[sloped, align=center, anchor=west, rotate=90] {5. pool threshold\\reached} (s6);
		
		\path[->] (s6) edge[bend left=10] node[sloped, align=center, anchor=east, rotate=270] {6. sanity check finished,\\no malicous\\behavior detected} (s2);
		
		\path[->] (s6) edge[bend right=0] node[sloped, align=center] {7. malicious behavior\\of party itself detected} (s7);
		
		\path[->] (s6) edge[bend right=23] node[sloped, align=center] {8. malicious coordinator\\detected} (s8);
		
		\path[->] (s4) edge[bend left=0] node[sloped, align=center, anchor=east, rotate=295.5, yshift=5, xshift=7] {9. malicious behavior\\of party itself\\detected}  (s7);
		
		\path[->] (s4) edge[bend left=0] node[sloped, align=center, anchor=west, rotate=57.9, yshift=5, xshift=-5] {10. malicious\\coordinator\\detected} (s8);
	
%		\draw (13, -1.56) -- (23, -1.56);

	\end{tikzpicture}	
\end{adjustbox}
	\caption{State chart showing the possible states of a party together with the events/actions that trigger state changes.}
	\label{fig:state_chart}
\end{figure*}
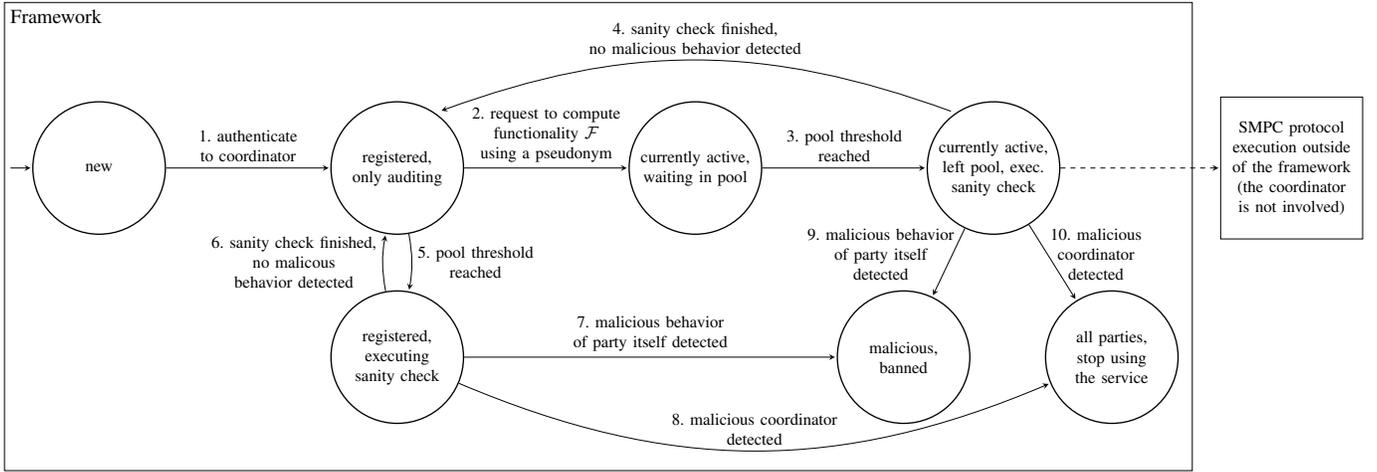	

\customParagraph{Parties:} There are two types of parties in our framework, i.e., the \emph{registered parties} and the \emph{currently active parties}. The set of registered parties includes all parties that successfully authenticated to the coordinator and are allowed to participate in executions of the SMPC protocol for computing~$ \func $. The currently active parties are those parties who have indicated their intent in participating in the next SMPC protocol execution for the computation of~$ \func $. They are anonymous within the set of registered parties which constitutes the anonymity set. To stay anonymous, the currently active parties use a pseudonym when indicating their intent to participate in a protocol execution. These pseudonyms are established between each party and the coordinator at registration such that it is infeasible (even for the coordinator) to link the pseudonym of a currently active party to the identity of the corresponding registered party. After participating in an SMPC protocol execution for the computation of~$ \func $, a party obtains a new pseudonym with the same properties which additionally cannot be linked to the party's previous pseudonym.

\customParagraph{Bulletin board:} Our \framework includes an append-only bulletin board where the coordinator publishes the necessary data to allow the registered parties to audit her behavior. The coordinator has read and write access to the bulletin board whereas the parties only have read access. 
The bulletin board consists of three components: the list of registered parties, the threshold pool, and the list of deprecated pseudonyms. 

The list of registered parties includes the identities (and the corresponding authentication tokens) of all parties that correctly authenticated to the coordinator and were admitted to participate in SMPC protocol executions.

The threshold pool includes the pseudonyms of the currently active parties (i.e., parties that currently intend to compute~$ \func $).

The list of deprecated pseudonyms is maintained to keep a record of all previously used pseudonyms, thus ensuring that each pseudonym can be added to the threshold pool exactly once (and thus be used exactly in one computation of~$ \func $).

To ensure that the coordinator behaves as intended, all registered parties have to audit the bulletin board each time it is updated by the coordinator and thus verify the validity of all information that is published on the bulletin board. In particular, they have to audit that:
(1) The authentication token of a party that is newly added to the list of registered parties is valid. (2) Each new pseudonym that is added to the threshold pool is valid and not deprecated. (3) All pseudonyms that were in the threshold pool are added to the list of deprecated pseudonyms. (4)~The coordinator removes a party from the list of registered parties if malicious behavior of that party has been detected. Note that an implementation of the framework has to ensure that the validity of the authentication tokens and the pseudonyms can be verified by any registered party. 

Similar to the voting system \emph{Helios}~\cite{Adida_Helios_2008}, we assume that the registered parties audit the bulletin board over time such that the chance is high that the coordinator is caught if she deviates from her specified behavior.
Our design assumes that if the parties notice that the coordinator does not behave as intended, they immediately stop using the service.

\subsection{Life Cycle}\label{sub:system_overview}

We describe the life cycle of a party within the \framework based on a state diagram (Figure~\ref{fig:state_chart}) that details the states that a party can assume together with possible state transitions. The life cycle of a party consists of four major parts.

\customParagraph{Registration:} If a party intends to participate in the computation of the functionality~$ \func $ provided by the framework, it first has to register with the coordinator (cf.~Transition~1). If the coordinator grants the party access to the service, she publishes the party's identity together with the authentication token on the bulletin board such that the other registered parties can audit this information. 
Furthermore, the coordinator and the newly registered party jointly establish a pseudonym for the party such that neither the coordinator nor the other registered parties can link the pseudonym to the party's identity. At the same time the coordinator and the other registered parties have to be able to verify that the pseudonym belongs to a registered party, that the pseudonym has not been used before, and that the party is reachable under the pseudonym.\footnote{Reachability under the pseudonym means that the coordinator and other registered parties can send messages to a party using the party's pseudonym without knowing the actual identity of the party behind the pseudonym.} Subsequently, the party then uses this pseudonym to remain anonymous when communicating with the coordinator or other registered parties. 

\customParagraph{Request to Compute the Functionality:} If a party intends to participate in the computation of the functionality~$ \func $, it indicates this to the coordinator using the previously established pseudonym (cf.~Transition~2). To prevent the coordinator from linking the pseudonym to the party's identity, the party has to use an anonymous communication channel when sending the request to compute~$ \func $ to the coordinator. The coordinator then verifies that the pseudonym is valid and not deprecated. If the request is valid, the coordinator updates the bulletin board accordingly, i.e., she adds the pseudonym to the threshold pool and to the list of deprecated pseudonyms.

\customParagraph{Sanity Check:} If the pool reaches its threshold, the coordinator removes all parties from the pool and all registered parties start the execution of a sanity check (cf.~Transitions~3~and~5). During the sanity check the parties have to gain assurance that: (1) The coordinator only added a pseudonym to the threshold pool for exactly those registered parties that indicated their intent to compute the functionality~$ \func $. (2) The coordinator only added pseudonyms of registered parties to the threshold pool. (3) The coordinator only added at most one pseudonym for each registered party to the threshold pool.

If the sanity check fails, all registered parties together with the coordinator determine whether the check failed because a party acted maliciously or because the coordinator acted maliciously. If malicious behavior of a party is detected, that particular party is removed by the coordinator, i.e., the coordinator removes the party from the set of registered parties and the party is denied any future participation in an SMPC protocol execution for computing the functionality~$ \func $ (cf.~Transitions~7~and~9). All other parties move to the state where they only audit the bulletin board (cf.~Transitions~4~and~6). If the check failed due to malicious behavior of the coordinator, all registered parties immediately stop using the service (cf.~Transitions~8~and~10). 

If the sanity check is successful (i.e., no malicious behavior of any party or the coordinator is detected), all parties return to auditing the bulletin board (cf.~Transitions~4~and~6). Each party that was in the pool establishes a new pseudonym with the coordinator with the same properties as the previous one. 
Additionally, it has to be impossible for any registered party or the coordinator to link the new pseudonym to the old one. 

\customParagraph{Computation of the Functionality:} The parties that were in the pool when the threshold was reached jointly execute the SMPC protocol for the computation of~$ \func $ (indicated by the dashed arrow in Figure~\ref{fig:state_chart}). During the protocol execution, the parties use their pseudonyms to communicate with each other such that their real identities remain private. Note that the actual protocol execution is not part of the framework, i.e., the coordinator does not participate in the protocol execution. Still, the protocol for computing~$ \func $ has to be secure against a dishonest majority as the \framework allows for the registration of new parties over time and it is thus not reasonable to assume an honest majority within the threshold pool.

\subsection{Threat Model and Assumptions}\label{sub:goals_threatmodel}

We assume that there is an adversary~$ \adversary $ who controls a set of corrupted parties~$ \corparties $ which forms a subset of all registered parties~$ \parties $ and/or the coordinator. The adversary can instruct the corrupted parties and/or the coordinator to arbitrarily deviate from their specified behavior. 
The adversary has access to all information that the corrupted parties obtain during their participation in the framework. This includes all information published on the bulletin board, all transcripts of interaction between the coordinator and the corrupted parties,  and all information the corrupted parties obtain during the sanity check execution and their participation in executions of the SMPC protocol for computing the functionality~$ \func $. If the adversary controls the coordinator, he also learns all transcripts of interaction between the honest parties and the coordinator.

However, it is considered impossible for the adversary to obtain any information that is only exchanged between honest parties. This includes the messages exchanged between honest parties during the execution of the SMPC protocol for computing the functionality~$ \func $ as well as the messages exchanged between the coordinator and the honest parties (given that the coordinator is not controlled by the adversary). Furthermore, we assume that the adversary is unable to forge messages of honest parties.

Finally, we assume that a party possesses an \emph{inalienable authentication token} and uses this token to authenticate to the coordinator at registration. An authentication token is called inalienable if it guarantees that the adversary can neither impersonate another party nor prevent a party from authenticating~\cite{Achenbach_InalienableAuthentication_2015}. This assumption is necessary to prevent the adversary from corrupting the service by registering fake parties. Furthermore, it allows the honest parties to connect malicious behavior to the identity of a malicious party which in turn allows for the permanent removal of that party. Note that we assume that a party has only one identity and is only able to obtain one inalienable authentication token.

\begin{figure*}[t!]
	\includegraphics[width = \textwidth]{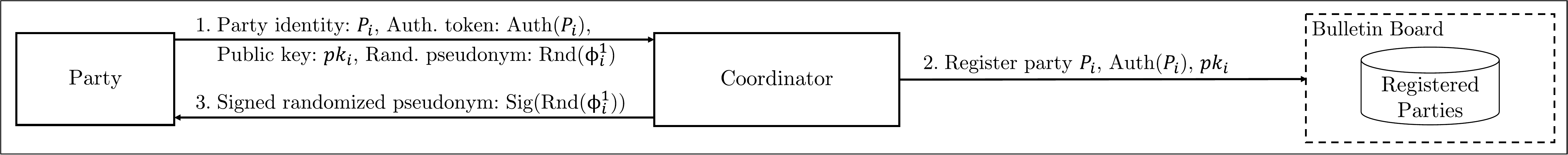}
	\caption{Registration of a party based on identity, inalienable authentication token, public key, and randomized pseudonym. 
	}
	\label{fig:registration}
\end{figure*}

\subsection{Security Properties}\label{sub:goals_secGoals}

We denote the index set of the registered parties $ \partysymbol{1}, ..., \partysymbol{\numparties} $ by $ \parties = \{1, ..., \numparties\} $. Additionally, with~$ \partysymbol{i} $ we also refer to the identity of the party. We denote the $ k $-th pseudonym of a party~$ \partysymbol{i} $ by $ \Pseudonym{i}{k} $. The transcript of interaction~$ \transcript{\Pseudonym{i}{k}} $ of a party~$ \partysymbol{i} $ denotes the complete interaction of the party~$ \partysymbol{i} $ with the coordinator under the pseudonym~$ \Pseudonym{i}{k} $. 
We identify the following four security properties which an implementation of our framework has to meet.

\begin{property}\label{def:party_anonymity}(Party Anonymity) 
	\textit{
		Assume that given the set~$ \parties $ of all registered parties and a party~$ \partysymbol{i} $'s ($ i \in \parties \setminus \corparties $) pseudonym~$ \Pseudonym{i}{k} $ that has been published on the bulletin board, an adversary~$ \adversary $ makes a guess $ X $ on the identity of a party. The probability $ \prob{X = \partysymbol{i}} $, i.e., the probability that the adversary~$ \adversary $ guesses the correct identity is $ \prob{X = \partysymbol{i}} = \frac{1}{\vert \parties \vert - \vert \corparties \vert} $ where $ \corparties $ corresponds to the index set of the parties controlled by the adversary.
	}
\end{property}

\begin{property}\label{def:party_unlinkability}(Party Unlinkability) 
	\textit{
		Given the pool threshold~$ \threshold $ and two pseudonyms~$ \Pseudonym{i}{k} $ and $ \Pseudonym{j}{l} $ that were not part of the same SMPC protocol execution with the corresponding transcripts of interaction $ \transcript{\Pseudonym{i}{k}} $ and $ \transcript{\Pseudonym{j}{l}} $, the adversary~$ \adversary $ makes a guess whether both pseudonyms belong to the same uncorrupted party. In the best case for the adversary, the pool both times included exactly the same set of parties and both times all corrupted parties were in the pool. In that case, the probability~$ \prob{\partysymbol{i} = \partysymbol{j}} $ that the adversary~$ \adversary $ correctly guesses that the two pseudonyms belong to the same party is $ \prob{\partysymbol{i} = \partysymbol{j}} = \frac{T - \vert \corparties \vert}{(T - \vert \corparties \vert)^2} =  \frac{1}{T - \vert \corparties \vert} $.
	}
\end{property}

Note that if the two pseudonyms were part of the same protocol execution, they could not belong to the same party as only one pseudonym per registered party is allowed inside the threshold pool at any given point in time. However, this does not influence the property that two pseudonyms can not be linked to one party.

\begin{property}\label{def:party_legitimacy}(Party Legitimacy) 
	\textit{
		For any party~$ \partysymbol{i} $ that is not a registered party ($ i \notin \parties $), it is computationally infeasible to obtain a valid pseudonym~$ \Pseudonym{i}{k} $ that the party could use to participate in an SMPC protocol execution for the computation of the functionality~$ \func $.
	}
\end{property}

\begin{property}\label{def:impersonation_resistance}(Impersonation Resistance)
	\textit{
		For any registered party~$ \partysymbol{i} $ ($ i \in \parties $), it is computationally infeasible to impersonate any other registered party~$ \partysymbol{j} $ ($ j \in \parties $ and $ i \neq j $).
	}
\end{property}

Note that the standard security property of SMPC protocols (i.e., a party learns nothing more than its private input and output) is not a property of our \framework but of the SMPC protocol that is used to compute the functionality~$ \func $. As the actual protocol execution is not part of our framework, this property has to be ensured by the SMPC protocol that is used. 

\section{Proposed Implementation}\label{sec:system_implementation}

In this section, we present a first implementation for the newly introduced \framework and discuss how it achieves the necessary security properties identified in Section~\ref{sub:goals_secGoals}.

\subsection{Implementation}\label{sub:system_specification}

\begin{figure*}[t!]
	\includegraphics[width = \textwidth]{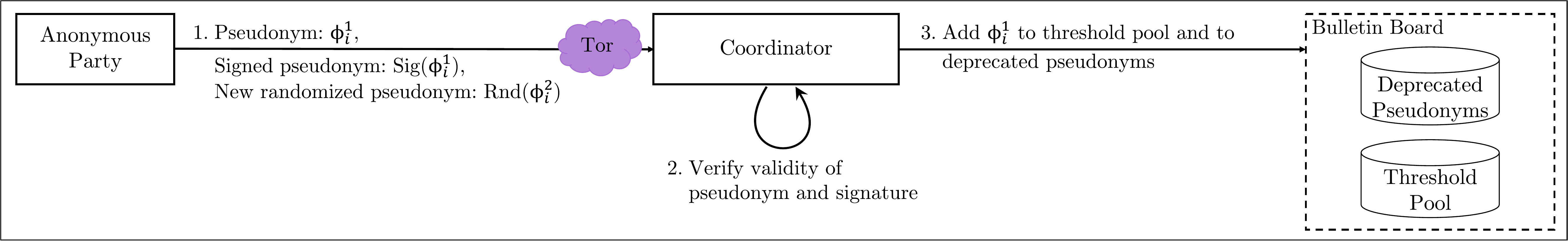}
	\caption{Request of an anonymous party to compute the functionality~$ \func $ containing the party's pseudonym, the signature on the pseudonym, and a new randomized pseudonym for future requests. }
	\label{fig:request}
\end{figure*}

\customParagraph{Registration:} 
A party $ \partysymbol{i} $ registers with the coordinator by providing its identity~$ \partysymbol{i} $, its inalienable authentication token~$ \auth{\partysymbol{i}} $, its public key~$ \pubkey{i} $, and a randomized pseudonym~$ \text{Rnd(\Pseudonym{i}{1})} $ to the coordinator (cf.~Transition~1,~Figure~\ref{fig:registration}). Recall that an inalienable authentication token guarantees that an adversary cannot impersonate a party using such a token~(cf.~Section~\ref{sub:goals_threatmodel}).

In our implementation, a party's pseudonym corresponds to the Tor onion service URL (cf.~Section~\ref{sub:pre_tor}) which the party uses to remain anonymous towards the coordinator when indicating its intent to compute the functionality~$ \func $. This makes it computationally infeasible to guess the pseudonym of a party since the onion URL corresponds to the hash of the public onion service identity~key~\cite{Tor_Onion_Service}. Furthermore, it allows the coordinator (and the other parties) to verify the validity of a party's pseudonym by checking that the party is reachable under the onion URL associated with the pseudonym. 

The coordinator checks the validity of the authentication token and verifies that the party is not already registered. Then, she adds the party's identity~$ \partysymbol{i} $ together with the public key $ \pubkey{i} $ and the authentication token~$ \auth{\partysymbol{i}} $ to the list of registered parties on the bulletin board (cf.~Transition~2,~Figure~\ref{fig:registration}). This allows all registered parties to also verify the validity of the authentication token. 
The public key~$ \pubkey{i} $ is used later on in our implementation of the sanity check procedure.

The coordinator then provides the party with a blind signature~$ \text{Sig}(\text{Rnd}(\Pseudonym{i}{1})) $ on the randomized pseudonym (cf.~Transition~3,~Figure~\ref{fig:registration}). 
A blind signature is a signature on a message that hides the content of the message from the signer by means of randomization~\cite{Chaum_BlindSignatures_1984}. After being signed, the owner can remove the randomization. We use blind signatures to enable a party to prove that it is a registered party without providing its real identity.
In particular, the blind signature scheme ensures that a party can only create exactly one valid pseudonym~$ \Pseudonym{i}{k} $ from a signed randomized pseudonym~$ \sig{\rnd{\Pseudonym{i}{k}}} $ and it makes it impossible to link the signed pseudonym to the party's identity. A possible candidate for such a scheme is the Okamoto-Schnorr blind~signature~scheme~\cite{Okamoto_BlindSignatures_1992}.

Upon receiving the signed randomized pseudonym $ \sig{\rnd{\Pseudonym{i}{1}}} $, the party removes the randomization and obtains the signed pseudonym~$ \text{Sig}(\Pseudonym{i}{1}) $. Thereby, the party obtains a pseudonym that cannot be linked to its identity as it is computationally infeasible to compute the pseudonym from the randomized pseudonym. At the same time, the signature on the pseudonym proves that the party is a registered party. 

\begin{figure*}[t]
	\begin{subfigure}{\textwidth}
		\includegraphics[width=\textwidth]{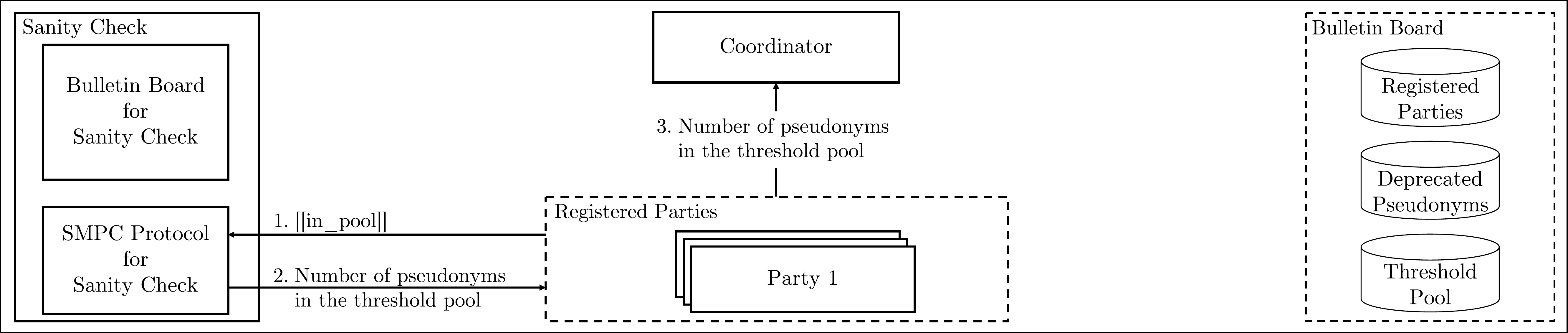}
		\subcaption{Before the anonymous parties in the threshold pool execute the SMPC protocol for the computation of the functionality~$ \func $, all registered parties jointly execute a sanity check to verify that the state of the threshold pool as indicated on the bulletin board is consistent with the set of registered parties that actually have a pseudonym within the pool.}
		\label{subfig:sanity_check}
	\end{subfigure}
	
	\vspace*{0.2cm}
	
	\begin{subfigure}{\textwidth}
		\includegraphics[width=\textwidth]{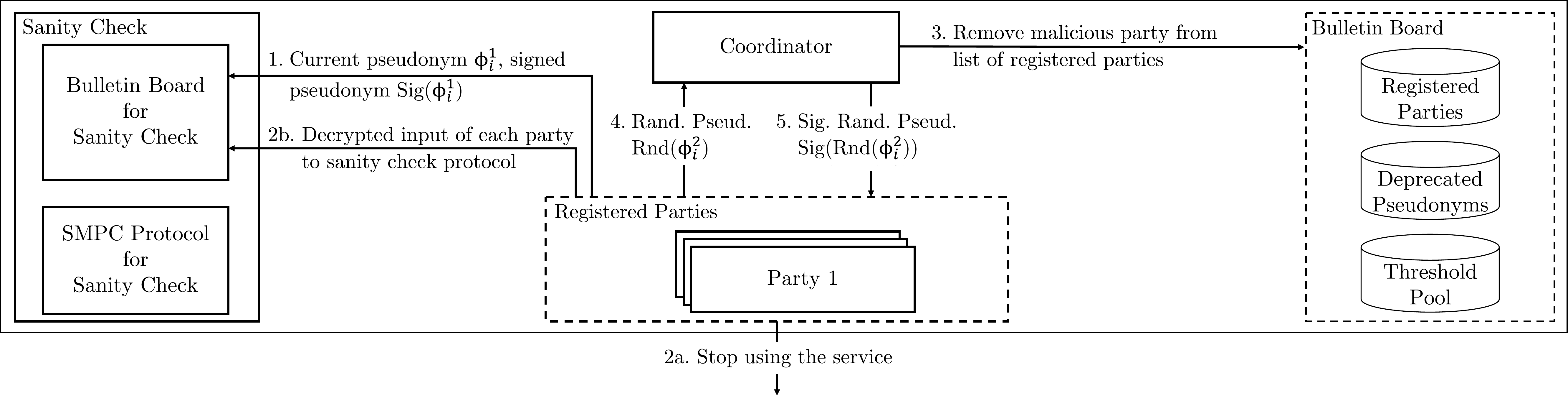}
		\subcaption{If the sanity check failed, the registered parties publish their current pseudonyms with the corresponding signatures on the bulletin board of the sanity check. If they thereby detect malicious behavior of the coordinator, they stop using the service. Otherwise, they jointly decrypt each party's input to the sanity check protocol and also publish it on the bulletin board. If malicious behavior of a party is detected, the coordinator removes the party from the list of registered parties. Afterwards, all remaining registered parties obtain new signed pseudonyms.}
		\label{subfig:sanity_check_failed}
	\end{subfigure}
	
	\vspace*{0.2cm}
	
	\begin{subfigure}{\textwidth}
		\includegraphics[width=\textwidth]{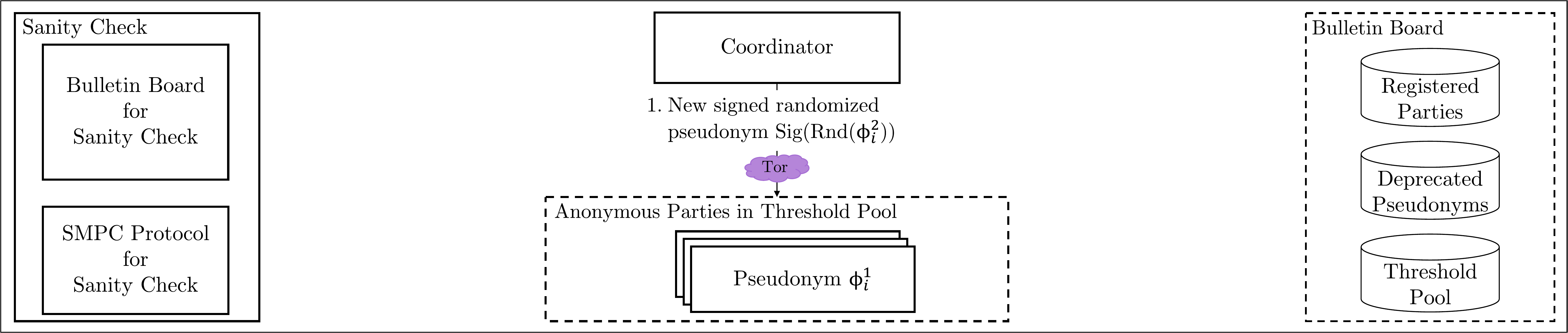}
		\subcaption{If the sanity check succeeds, the coordinator provides the parties that were in threshold pool with blind signatures on new randomized pseudonyms using the old pseudonyms (onion URLs)) as receiver addresses as she does not know the identities of the pseudonyms in the threshold pool.}
		\label{subfig:sanity_check_success}
	\end{subfigure}
	
	\caption{Sanity check executed among all registered parties and the coordinator to verify the integrity of the threshold pool.}
\end{figure*}

\customParagraph{Request to Compute the Functionality:} 
If a registered party~$ \partysymbol{i} $ ($ i \in \parties $) intends to participate in the computation of the functionality~$ \func $, it sends a request to the coordinator over the Tor network. This prevents the coordinator from linking the request to a party's identity. The request contains the party's current pseudonym~$ \Pseudonym{i}{1} $, the signature on that pseudonym~$ \text{Sig}(\Pseudonym{i}{1}) $, and a new randomized pseudonym~$ \text{Rnd}(\Pseudonym{i}{2}) $ (cf.~Transition~1,~Figure~\ref{fig:request}). 

The coordinator verifies that the party is a registered party by checking the correctness of the signature. Furthermore, the coordinator has to verify that the party is reachable under the indicated pseudonym. To this end, she can just send a request to the corresponding onion URL. If the party sends back a response, the coordinator knows that the party indeed owns the onion URL associated with its pseudonym. 
Finally, the coordinator verifies that the pseudonym has not been used before by checking that the pseudonym is not on the list of deprecated pseudonyms (maintained on the bulletin board). 

If the pseudonym is valid and not deprecated and the signature is valid, the coordinator adds the party to the threshold pool, i.e., she adds the pseudonym and the signature on it to the pool.
Furthermore, the coordinator adds the pseudonym to the list of deprecated pseudonyms such that it cannot be used twice by the party (cf.~Transition~3,~Figure~\ref{fig:request}).
Recall that both the threshold pool and the list of deprecated pseudonyms are maintained on the bulletin board and that all registered parties have to audit the validity of this information.

\customParagraph{Sanity Check - Procedure:} When the pool reaches its threshold, all~registered parties start the execution of a sanity check. As stated in Section~\ref{sub:system_overview}, this check has to ensure (1) that the coordinator added exactly those registered parties to the pool that indicated their intent to compute the functionality~$ \func $, (2)~that the coordinator only added pseudonyms of registered parties to the pool, and (3)~that the coordinator added at most one pseudonym for each registered party to the pool. 

The first property is achieved by the blind signature scheme. As the coordinator is unable to obtain the pseudonym of a party only from the randomized pseudonym which the party sent to the coordinator at registration, she has no means to obtain a valid pseudonym of a party. Furthermore, we assume that even a corrupted coordinator does not act maliciously at random. This implies that the coordinator will not deny a party to participate in the computation of~$ \func $ if the party sends a request containing a valid and correctly signed pseudonym that is not deprecated as the coordinator does not know the actual party identity belonging to the pseudonym. Hence, the coordinator adds the pseudonyms of all registered parties that indicated their intent to compute~$ \func $ to the threshold pool.

Our sanity check thus only has to ensure that properties (2) and (3) hold. To this end, all registered parties execute an SMPC protocol that computes the number of registered parties that currently have a pseudonym inside the pool. Thereby, it is guaranteed that the the bulletin board is consistent with the actual state of all registered parties. 

Such a protocol can be implemented using an \emph{additive homomorphic threshold encryption scheme} (e.g.,\cite{FouquePaillierThreshold2001}) and \emph{efficient zero-knowledge range proofs} (e.g.,~\cite{Lipmaa_RangeProofs_2003}). Each party broadcasts an encrypted bit indicating whether the party currently has a pseudonym within the pool or not together with a range proof that the encrypted value is in the set $ \{0,1\} $. Then, each party can locally sum up all encrypted bits. Finally, the parties jointly decrypt the result and thus obtain the number of parties that currently have a pseudonym within the threshold pool (cf.~Transition~2, Figure~\ref{subfig:sanity_check}). Each party then sends this number to the coordinator (cf.~Transition~3, Figure~\ref{subfig:sanity_check}). 

Note that this procedure has one drawback, i.e., if two (or more) corrupted parties collaborate, one party can claim that it has a pseudonym in the pool although it has not whereas a second party has two pseudonyms inside the pool. This behavior would not be detected and the sanity check would succeed. Although this violates the property that each party may only have one pseudonym inside the pool, it is equivalent to the case where one registered party participates in the protocol on behalf of another registered party which is always possible. In particular, it does not increase the maximum number of pseudonyms controlled by the adversary.

\customParagraph{Sanity Check - Failure:} If the number of pseudonyms inside the pool shown on the bulletin board does not match the number computed during the SMPC protocol for sanity check, either a party or the coordinator acted maliciously. To detect the source of malicious behavior, we have to determine which parties currently have a pseudonym in the pool. To this end, each registered party has to publish its current pseudonym and the signature on it. However, the bulletin board operated by the coordinator cannot be used for this purpose as the parties only have read access and a corrupted coordinator cannot be trusted to correctly publish the data. Thus, we introduce a second bulletin board to which all registered parties have read and write access while the coordinator only has read access. This second bulletin board is specific to our implementation of the sanity check. It is not to be confused with the bulletin board of the framework (where the coordinator publishes her actions). To guarantee the authenticity of the data published on this second bulletin board, the parties use their private key $ \privkey{i} $ corresponding to the public key $ \pubkey{i} $ published by the coordinator during registration (cf.~Transition~2, Figure~\ref{fig:registration}).

The parties then publish their current pseudonyms and the signatures on them on this bulletin board (cf.~Transition~1, Figure~\ref{subfig:sanity_check_failed}). Upon this, all parties verify that the signatures on all published pseudonyms are valid. There are three possible reasons why the sanity check may fail:

(1) The signature on a party's pseudonym is not valid. Thus, the party acted maliciously and is banned by the coordinator (cf.~Transition~3,~Figure~\ref{subfig:sanity_check_failed}). Note that if the coordinator provided a party with an invalid signature, the party should have detected this and should not have used the pseudonym. 

(2) There is a pseudonym with a valid signature in~the threshold pool that does not correspond to the current pseudonym of any party. Thus, the coordinator either signed multiple pseudonyms for one party or signed a pseudonym for a party that is not registered. In both cases, malicious behavior of the coordinator is detected and all registered parties stop using the service (cf.~Transition~2a, Figure~\ref{subfig:sanity_check_failed}). 

(3) A party lied during the sanity check protocol. In this case, all pseudonyms in the threshold pool are among the currently valid pseudonyms of the parties, i.e., the state of the pool as published on the bulletin board is correct. To determine the party that lied during the sanity check, all parties jointly decrypt each party's input to the sanity check protocol and publish it on the bulletin board of the sanity check (cf.~Transition~2b, Figure~\ref{subfig:sanity_check_failed}). The parties and the coordinator can then compare each party's current pseudonym and the input bits used in the sanity check protocol. If there is a party whose input bit does not match its state on the bulletin board, the party lied during the sanity check. This party can then be banned by the coordinator (cf.~Transition~3,~Figure~\ref{subfig:sanity_check_failed}). 

After the exclusion of a party, each remaining registered party sends a new randomized pseudonym to the coordinator who returns a blind signature on it such that afterwards all parties have validly signed pseudonyms (that are not deprecated) which can neither be linked to their actual identities nor to their previous ones (cf.~Transitions~4~and~5, Figure~\ref{subfig:sanity_check_failed}). To guarantee the authenticity of this process, the parties sign their request for a new pseudonym (cf.~Transition~4, Figure~\ref{subfig:sanity_check_failed}) with their private key~$ \privkey{i} $ corresponding to the public key~$ \pubkey{i} $ sent to the coordinator at registration.

Note that our implementation of the sanity check implies that the anonymity of the parties is compromised if the sanity check fails. However, this is necessary to be able to track down the malicious behavior to one specific party and thus be able to remove that party. As each time the sanity check fails, at least one corrupted party is banned by the coordinator, the adversary can only force the de-anonymization of the parties for a limited number of times. Furthermore, the de-anonymization during the sanity check only compromises anonymity for that specific request to compute the functionality~$ \func $. For all past and future requests of a party, anonymity is still guaranteed.

\customParagraph{Sanity Check - Success:} If the sanity check is successful (the number of pseudonyms in the threshold pool indicated on the bulletin board matches the number computed during the sanity check protocol), the coordinator provides each party~$ \partysymbol{i} $ that was in the threshold pool with a signature~$ \text{Sig}(\text{Rnd}(\Pseudonym{i}{2})) $ on their new randomized pseudonym which the parties sent to the coordinator along with their request to compute the functionality~$ \func $ (cf.~Transition~1, Figure~\ref{subfig:sanity_check_success}). Thus, the parties only receive a new signed pseudonym after they have left the threshold pool (and completed the sanity check). Thereby, it is guaranteed that each party can only have one pseudonym in the pool at each point in time. 
The parties can then use their new pseudonyms for future requests to compute~$ \func $.

\customParagraph{Computation of the Functionality:} 
If the sanity check succeeds, the parties that were in the threshold pool start the execution of an SMPC protocol for the computation of~$ \func $.~They use their pseudonyms (onion URLs) as their addresses for the communication during the SMPC protocol. Thereby, they stay anonymous to each other during the protocol execution.

\subsection{Discussion}\label{sub:system_discussion}

In this section, we analyze to what extent our implementation of the \framework outlined in Section~\ref{sub:system_specification} fulfills the security properties defined in Section~\ref{sub:goals_secGoals}.

\customParagraph{Party anonymity:}
A party provides its identity to the coordinator at registration and the coordinator adds it to the list of registered parties on the bulletin board (cf.~Figure~\ref{fig:registration}). Thus, the set of registered parties $ \parties $ forms the anonymity set. In addition, the party receives a blind signature~$ \text{Sig}(\text{Rnd}(\Pseudonym{i}{1})) $ on its randomized pseudonym~$ \text{Rnd}(\Pseudonym{i}{1}) $. The first time a party sends a request to compute functionality~$ \func $, the party uses its pseudonym $ \Pseudonym{i}{1} $ as its identifier and the signature on the pseudonym to prove that it is a registered party (cf.~Figure~\ref{fig:request}). As the party only sends its randomized pseudonym to the coordinator at registration and as the party uses the Tor network when sending the request, the coordinator is not able to link the pseudonym to a registered party. The only information that the coordinator obtains is that the pseudonym belongs to one of the registered parties (as only these parties are able to obtain a correct signature on their pseudonym). Thus, the chance that the coordinator correctly guesses the party identity belonging to the pseudonym is $ \frac{1}{\vert \parties \vert} $.

As the coordinator publishes each pseudonym that is added to the threshold pool on the bulletin board, all other parties also learn the pseudonym. The chance that a registered party~$ \partysymbol{j} $ correctly guesses that a pseudonym~$ \Pseudonym{i}{k} $ belongs to another party $ \partysymbol{i} $ with $ i \neq j $ is $ \frac{1}{\vert \parties - 1 \vert} $ (party $ \partysymbol{j} $ can exclude that the pseudonym belongs to itself).

An adversary~$ \adversary $ who has corrupted~$ \vert \corparties \vert $ parties can exclude that a pseudonym $ \Pseudonym{i}{1} $ belongs to any of the corrupted parties controlled by himself. Thus, the probability that the adversary correctly guesses the party $ \partysymbol{i} $ belonging to the pseudonym~$ \Pseudonym{i}{1} $ is $ \frac{1}{\vert \parties \vert - \vert \corparties \vert} $ for $ i \notin \corparties $. If $ i \in \corparties $, the adversary knows that the pseudonym belongs to party $ \partysymbol{i} $.
The same holds for any new pseudonym $ \Pseudonym{i}{k} $ which a party obtains after a successful sanity check since this new pseudonym is again only sent to the coordinator in randomized form and using the Tor network. 

The only situation in which the above does not hold is if the sanity check protocol yields a different number of pseudonyms inside the threshold pool than indicated on the bulletin board. In that case, the current pseudonym of each registered party is published and thus the identities of those parties that are currently in the threshold pool are leaked. However, this is necessary to enable the detection of the corrupted party or coordinator that caused the sanity check failure. Furthermore, the adversary cannot force this situation infinitely often as each time the identity of a corrupted party is leaked, that party is banned. Also the anonymity of the parties is only compromised for that specific request to compute the functionality. In particular, publishing their pseudonyms for this one request does not influence party anonymity w.r.t.\ previous or future requests since our implementation provides for party unlinkability (cf.~\secprop~\ref{def:party_unlinkability}).

Thus, our implementation provides for party anonymity (cf.~\secprop~\ref{def:party_anonymity}) as long as the sanity check succeeds.

\customParagraph{Party unlinkability:}
In order to obtain a valid signature on a new pseudonym~$ \Pseudonym{i}{l} $, a party $ \partysymbol{i} $ sends the randomized pseudonym~$ \text{Rnd}(
(\Pseudonym{i}{l})) $ to the coordinator when indicating its intent to compute the functionality~$ \func $ using its current pseudonym~$ \Pseudonym{i}{k} $  (cf.~Figure~\ref{fig:request}). After the successful sanity check on the threshold pool containing the pseudonym~$ \Pseudonym{i}{k} $, the coordinator uses the old pseudonym~$ \Pseudonym{i}{k} $ (i.e., the Tor onion service URL) to provide party~$ \partysymbol{i} $ with a signature on the new pseudonym~$ \Pseudonym{i}{l} $. As it is computationally infeasible to derive $ \Pseudonym{i}{l} $ from $ \text{Rnd}(\Pseudonym{i}{l}) $, it is also computationally infeasible for the coordinator to establish a link between $ \Pseudonym{i}{k} $ and~$ \Pseudonym{i}{l} $. 

A registered party does not obtain any information on a pseudonym $ \Pseudonym{i}{k} $ that is published on the bulletin board but that it belongs to another registered party. Thus, it is also computationally infeasible for a party to correctly guess that $ i = j $ given two pseudonyms $ \Pseudonym{i}{k} $ and $ \Pseudonym{j}{l} $ and the corresponding transcripts of actions $ \transcript{\Pseudonym{i}{k}} $, $ \transcript{\Pseudonym{j}{l}} $. 
In particular, assuming that the adversary~$ \adversary $ controls the set of corrupted parties~$ \corparties $ and that these are always in the pool of threshold $ T $, the probability~$ \prob{\partysymbol{i} = \partysymbol{j}} $ that the adversary correctly guesses that $ i = j $ for two pseudonyms $ \Pseudonym{i}{k} $ and $ \Pseudonym{j}{l} $ ($ i, j \notin \corparties $) that were not part of the same protocol execution is~$ \frac{1}{T - \vert \corparties \vert} $. 

The only situation in which the adversary learns anything about the relation between two pseudonyms $ \Pseudonym{i}{k} $ and $ \Pseudonym{j}{l} $ is when observing that the two pseudonyms are part of the same protocol execution. However, in this case the adversary only learns that the two pseudonyms do not belong to the same party (i.e., $ i \neq j $). This does not allow the adversary to link different pseudonyms to the same party.

Thus, our implementation provides for party unlinkability as stated in \secprop~\ref{def:party_unlinkability}.

\customParagraph{Party legitimacy:}
If a party sends a request to compute the functionality~$ \func $, the coordinator verifies that the party belongs to the set of registered parties by checking the validity of the presented signed pseudonym. Furthermore, the coordinator makes sure that the pseudonym has not been used before by comparing it to a list of deprecated pseudonyms (cf.~Figure~\ref{fig:request}). 

If the adversary controls the coordinator, he can of course make the coordinator sign a pseudonym that does not belong to any registered party or add a pseudonym to the threshold pool although the signature is not valid. However, this would be detected by the registered parties during the sanity check protocol which would yield a different number of pseudonyms in the threshold pool than shown on the bulletin board. In that case the parties would publish their current pseudonyms and notice that there is a pseudonym that is not correctly signed or one that is correctly signed but that does not belong to any registered party. In both cases, they would know that the coordinator is corrupted and they would stop using the service.

There is only one possibility for a party $ \partysymbol{i} $ ($ i \notin \parties $) to have a pseudonym in the threshold pool although the party is not a registered party. If the adversary controls the coordinator and at least one registered party $ \partysymbol{j} $ ($ j \in \parties $), he can make the coordinator sign a pseudonym for the party $ \partysymbol{i} $ although the party is not a registered party. During the sanity check the corrupted registered party $ \partysymbol{j} $ (which does not have a pseudonym in the threshold pool) then falsely indicates that it has a pseudonym inside the pool. This would not be noticed during the sanity check. However, this scenario is equivalent to making a corrupted registered party enter the pool with the input of an unregistered party which cannot be prevented in our implementation. In particular, this does not increase the number of pseudonyms of corrupted parties that can be inside the threshold pool.

Thus, our implementation provides for party legitimacy as stated in \secprop~\ref{def:party_legitimacy}.

\customParagraph{Impersonation resistance:}
At registration a party has to provide the coordinator with an inalienable authentication token~\cite{Achenbach_InalienableAuthentication_2015} that proves its identity. We assume that it is impossible to forge or steal such a token in order to impersonate a party. Consequently, a party controlled by the adversary is not able to register using the identity of another party. 

Thus, we only have to show that it is impossible for a party to use the pseudonym of another party when sending a request to the coordinator. However, this is trivial as a party has no means to obtain the pseudonym of another party before that pseudonym is published on the bulletin board since the first time a party sends its pseudonym in plain is exactly when indicating its intent to compute the functionality~$ \func $ (cf.~Figure~\ref{fig:request}). After the publication of the pseudonym on the bulletin board, all parties and the coordinator consider the pseudonym deprecated. Thus, they would not accept a second request using that pseudonym. 

Note that it is considered computationally infeasible to correctly guess a party's pseudonym as the pseudonym corresponds to the Tor onion service URL of a party which is not public prior to its publication on the bulletin board. Onion service URLs are random values, i.e., they are computed as the hash of the public onion service identity key~\cite{Tor_Onion_Service}. Thus, the probability to correctly guess a party's pseudonym before being published on the bulletin board is negligible. This holds even if the adversary controls the coordinator as the coordinator only obtains a party's randomized pseudonym at registration and as it is computationally infeasible to deduce the pseudonym from the randomized pseudonym. 

Thus, our implementation provides for impersonation resistance as stated in \secprop~\ref{def:impersonation_resistance}.

\section{Conclusion and Future Work}\label{sec:conclusion}

In this paper, we have presented the first framework that allows a set of parties to remain anonymous during the repeated execution of an SMPC protocol for the computation of a functionality~$ \func $. At the same time our approach guarantees that only authorized parties may participate in the protocol executions. Besides specifying such a framework and the security properties that an implementation of the framework has to achieve, we have also introduced one possible implementation of the framework and shown in how far it achieves the previously defined security properties.

While our proposed implementation allows the parties to compute any functionality~$ \func $, implementations for special use cases may be developed in future work. These may then come with special additional requirements. For example, in our implementation the parties' anonymity is compromised if the sanity check fails such that the source of malicious behavior may be tracked down. However, there may be use cases where this is not acceptable. For such use cases, one would have to implement a different solution for connecting malicious behavior to a specific party. Furthermore, our framework currently does not include the possibility for a party to de-register. However, there may be use cases where it is not possible for all parties to always participate in the sanity check execution each time the functionality~$ \func $ has to be computed. For such use cases, it may then be necessary to add the possibility of de-registration such that those parties that are not able to always participate in the sanity check can de-register after their participation in the computation of~$ \func $. 

Thus, in general the implementation proposed in this paper may have to be extended or adapted in the future to fit the special requirements of different use cases.

\bibliographystyle{IEEEtranS}
\bibliography{references}

\end{document}